\documentclass[aps,twocolumn,showkeys,unsortedaddress]{revtex4-1}
\usepackage{graphicx}
\usepackage{amsmath}
\usepackage{amssymb}
\usepackage{longtable}
\usepackage{dcolumn}
\usepackage{natbib}
\begin{document}

\title{Examining the meaning of the peptide transfer free energy obtained from blocked (Gly)$_n$ and cyclic-diglycine model
compounds}
\author{D. Asthagiri}\thanks{dilipa@jhu.edu}
\author{Dheeraj S. Tomar}
 \affiliation{Department of Chemical and Biomolecular Engineering, Johns Hopkins University, Baltimore, MD 21218}
\author{Val{\'e}ry Weber}
 \affiliation{IBM Zurich Research}
\begin{abstract}
In experiments, the free energy of transferring the peptide group from water to an osmolyte solution is obtained using the 
transfer free energy of (Gly)$_n$ with the added assumption that a constant incremental change in free energy with $n$ 
implies that each additional unit makes an independent contribution to the free energy.  Here we test this assumption and
uncover its limitations. Together with results for cyclic-diglycine, we show that, in principle, it is not possible to obtain a peptide group transfer free 
energy that is independent of the model system.  We calculate the hydration free energy, $\mu^{\rm ex}_n$, of 
acetyl-(Gly)$_n$-methyl amide ($n=1\ldots7$) peptides modeled in the extended conformation in water and osmolyte solutions.   
$\mu^{\rm ex}_n$ versus $n$ is linear,  suggestive of independent, additive group-contributions. To probe the observed linearity further, 
we study the hydration of the solute bereft of water molecules in the first hydration shell. This conditioned solute arises naturally in the 
theoretical formulation and helps us focus on hydration effects uncluttered by the complexities of  short-range solute-water interactions. 
We subdivide the conditioned solute into $n+1$ peptide groups and a methyl end group. The binding energy of each of these groups 
with the solvent is Gaussian distributed,  but the near neighbor binding energies are themselves correlated: the $i,i+1$ correlation is the strongest 
and tends to lower the free energy over the independent group case.  We show that the observed linearity can be explained
by the similarity of near neighbor correlations. Implications for group additive transfer free energy models are indicated. 
\end{abstract}
\maketitle

\section{Introduction}

Group additive decomposition of the free energy of protein conformational change has a rich history in attempts to understand the 
physical factors governing protein stability in solution \cite{makhatadze:1995}.  Such efforts are at the heart of past and current efforts to understand how 
the solvent modulates protein folding thermodynamics \cite{Nozaki:1963,Tanford:1970rev,Auton:pnas05,Auton:pnas07,Auton:bc11}. Since the peptide bond 
is the most numerous group in a protein, attempts to obtain the transfer free energy of the peptide group have occupied a particularly important position in the 
broader attempt to understand the role of the solution in protein folding thermodynamics \cite{Nozaki:1963,Auton:bio04}. 
Indeed, such group additive transfer free energy analysis has been instrumental in revealing that conformation-protecting osmolytes primarily exert their influence by changing the solubility of the peptide backbone \cite{Bolen:jmb01,Bolen:rev08}, an identification with significant consequences to our understanding of protein folding \cite{Rose:pnasbackbone}. 
However, a clear theoretical analysis of the meaning of  the transfer free energy of the peptide group that apparently obeys group-additivity has not been satisfactorily established. Here we address this issue on the basis of a physically transparent theoretical framework and computer simulations. However, the insights from this work are not limited to a peptide group, but apply more broadly to all such group-additive decompositions of free energy that are used in studying protein folding and stability.

In seeking the contribution of the peptide, starting from the seminal work of Nozaki and Tanford \cite{Nozaki:1963}, it is common practice to consider the 
transfer free energy (typically from water to an aqueous solution of an additive) of glycyl-peptides of increasing chain length. The peptides
can be blocked n-acetylglycinamides (as in this study) or zwitterionic (as in the studies by Nozaki and Tanford \cite{Nozaki:1963}). The transfer free energy of the peptide
group then has been sought by considering the difference in transfer free energy of chains of length $m$ and $n$ ($m > n$) 
by various constructs; for example, for $m=n+1$, the peptide free energy would be equated with the free energy difference between 
the chains of length $m$ and $n$. A somewhat more robust approach termed the constant increment method equates the peptide transfer free energy to 
the slope of the transfer free energy with respect to $n$. Various such constructs are possible and these have been well-documented by Auton and Bolen \cite{Auton:bio04}.

Work by Auton and Bolen \cite{Auton:bio04} has also helped clarify some of the vexing issues related to the choice of concentration scales and model compounds in 
determining the peptide group transfer free energy \cite{Auton:bio04}.  By careful consideration of peptide solubility issues, these authors have showed that 
reasonably concordant values of the peptide transfer free energy can be obtained that are independent of the concentration scale
and of the model system --- glycyl-peptides versus cGG, the cyclic-diglycine molecule.  (For cGG, he peptide transfer free energy is sought by 
dividing the experimental transfer free energies by 2.) While this concordance is pleasing, it is also somewhat
troubling since in cGG the CO and NH of the peptide are \textit{cis} and the molecule has a net zero dipole moment, whereas in all usual proteins the CO and NH are
\textit{trans} and the glycyl peptides used in the experiments can have non-negligible dipole moment. Further, the $\phi, \psi$ angles in cGG are also not consistent with what is 
found in proteins (George Rose, personal communication).  Thus, either the conformation of the peptide is unimportant in the transfer free energy
value or there are other effects that lead to this result or a combination of both. 

Concerns about group additivity \cite{Dill:1997tg}, and in particular, identifying a group additive contribution for the peptide \cite{baldwin:proteins06,baldwin:pnas09}, are not new. 
Using a continuum dielectric model of the solvent, Avbelj and Baldwin  \cite{baldwin:proteins06,baldwin:pnas09} have argued that failure of group additivity arises due to
dependence of the hydration free energy of the peptide on the neighboring groups (that serve to occlude the solvent medium in their model). Our point complements this, but is
more broader. We show that an independent group additive contribution is not a consequence even when the conditions for use of 
the constant increment approach are satisfied. Moreover, both electrostatics and  van~der~Waals (dispersion) interactions  contribute to the failure of independence, each
in rather subtle ways. Thus care is needed even in decoupling free energy contributions of nonpolar groups from adjacent polar groups, an issue 
that had been anticipated before Ref.\ \cite{paulaitis:corr10}. 

Here we use theory and computer simulations to examine the vacuum to solution ($S$) transfer free energy of Acetyl-(Gly)$_n$-methyl amide peptides and 
of cGG. The free energies are obtained by a quasichemical organization of the potential distribution theorem. A virtue of this 
formulation is that is that it makes transparent the role of correlated fluctuations of the binding energies of two groups on the molecule and its role in the 
thermodynamics of hydration. A central observation of our work is that even for an idealized solute with no complicated short-range solute-solvent
interaction, the group-solvent binding energies between neighboring groups are correlated. This implies that identifying a group-contribution to free energy solely due 
to an individual group is, in principle, not possible, even for this idealized solute. The situation for a real solute is expected to be considerably more complicated. 

\section{Theory}

The excess chemical potential, $\mu^{\rm ex}$, of a solute in the solution is that part of the Gibbs free energy that would vanish if the interaction between the solute and 
solvent were to vanish. Formally,
\begin{eqnarray}
\beta\mu^{\rm ex} = \ln \int e^{\beta \varepsilon} P(\varepsilon) d\varepsilon \, ,
\label{eq:pdt}
\end{eqnarray}
where $\varepsilon = U_{S+1} - U_{S} - U_{\rm p}$ is the binding energy of the solute with the rest of the fluid. $U_{S+1}$ is the potential energy of the 
solute plus solvent system at a particular configuration of the solvent (we assume the peptide conformation to be fixed), 
$U_{S}$ is the potential energy of the same configuration but with the solute removed, and $U_{\rm p}$ is the potential energy of the solute, here the peptide, 
solely. $P(\varepsilon)$ is the probability density distribution of $\varepsilon$. $\mu^{\rm ex}$ is the
excess free energy in the liquid relative to an ideal gas at the same density and temperature. As usual, $\beta = 1/k_{\rm B}T$, where
$T$ is the temperature and $k_{\rm B}$ the Boltzmann constant. 

\begin{figure}[h!]
\includegraphics[width=3.0in]{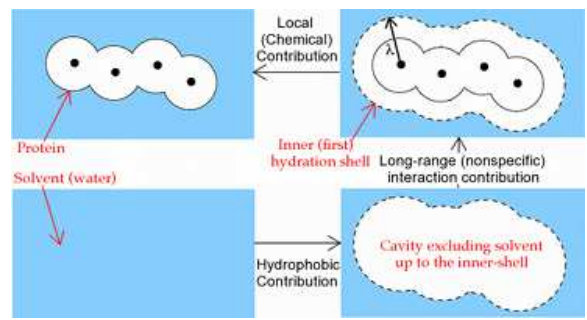}
\caption{Schematic showing the physical pieces contributing to the solvation free energy (Eq.~\ref{eq:qc}) of the protein. This decomposition follows the regularization of the
solute-solvent binding energy distribution.}
\label{fg:cycle}
\end{figure}

Following earlier work,  to calculate $\mu^{\rm ex}$ from Eq.~\ref{eq:pdt}, we regularize $P(\varepsilon)$ by introducing an auxiliary constraint,
a field $\phi_\lambda$ that pushes the solvent molecules away from the solute's surface to a range $\lambda$. This construct has the virtue of tempering the solute-solvent
 interaction, and, for solvent pushed far enough (typically evacuating the first hydration shell is sufficient), the distribution of binding energies is a Gaussian. Formally, 
with the introduction of the field, 
\begin{eqnarray}
\beta\mu^{\rm ex} = \underbrace{\ln x_0[\phi_\lambda]}_{\rm local\; chemistry} \underbrace{- \ln p_0[\phi_\lambda]}_{\rm packing} + \underbrace{\beta\mu^{\rm ex}[P(\varepsilon|\phi_\lambda)]}_{\rm long-range} \; .
\label{eq:qc}
\end{eqnarray}
$-\ln x_0[\phi_\lambda]$ is the free energy required to apply the field in the solute-solvent system: it reflects the strength of the solute interaction with the solvent in the
inner shell. $-\ln p[\phi_\lambda]$ is the free energy required to apply the field in the neat solvent system: it reflects the intrinsic properties of the solvent. For $\phi_\lambda$
modeling a hard exclusion of solvent,  $-\ln p_0[\phi_\lambda]$ is precisely the hydrophobic contribution to hydration \cite{Pratt:1992p3019,Pratt:2002p3001}. $\beta\mu^{\rm ex}[P(\varepsilon|\phi_\lambda)]$ is the contribution to $\beta\mu^{\rm ex}$ from long-range solute-solvent interactions. In molecular dynamics simulations, we calculate $-\ln x_0[\phi_\lambda]$ or $-\ln p_0[\phi_\lambda]$ simply by the work required to apply $\phi_\lambda$ \cite{weber:jctc12}. Fig.~\ref{fg:cycle} gives a schematic of the decomposition of $\mu^{\rm ex}$ 
according to Eq.~\ref{eq:qc}.

For $\phi_\lambda$ excluding solvent from the first hydration (or inner-shell), the conditional binding energy distribution $P(\varepsilon|\phi_\lambda)$ can be
well-described by a Gaussian of mean $\langle \varepsilon | \phi_\lambda\rangle$ and variance $\langle \delta\varepsilon^2 | \phi_\lambda\rangle$ \cite{weber:jcp11,merchant:jcp11a,asthagiri:jacs07,shah:jcp07} and the
long-range contribution is then given by 
\begin{eqnarray}
\mu^{\rm ex}[P(\varepsilon|\phi_\lambda)] = \langle \varepsilon | \phi_\lambda\rangle + \frac{\beta}{2} \langle \delta\varepsilon^2 | \phi_\lambda\rangle \, .
\label{eq:gaussian}
\end{eqnarray}

Now consider decomposing $\varepsilon$ into contributions $\varepsilon_i$ due to various groups $i = 1, \ldots, n$ comprising the solute under 
consideration. For a pairwise additive forcefield, as is used in this study, such a decomposition can be unambiguously made. For the conditioned solute, 
even the individual binding energy distributions $P(\varepsilon_i | \phi_\lambda)$ are Gaussian distributed, but, in general, $\varepsilon_i$ is correlated 
with $\varepsilon_j$ ($j \neq i$). In this case, $P(\varepsilon = \sum_i \varepsilon_i)$ is Gaussian
distributed  with a mean $ \sum_i \langle \varepsilon_i | \phi_\lambda\rangle$  and a variance 
$  \sum_i \langle \delta\varepsilon_i^2 | \phi_\lambda\rangle + 2 \sum_{i>j} \rho_{ij} \sqrt{\langle \delta\varepsilon_i^2 | \phi_\lambda\rangle \langle \delta\varepsilon_j^2 | \phi_\lambda\rangle}$, where $\rho_{ij}$ is the correlation coefficient \cite{paulaitis:corr10}.  The long-range contribution to the free energy  is then given by 
\begin{eqnarray}
\mu^{\rm ex}[P(\varepsilon|\phi_\lambda)] & = & \sum_i \mu^{\rm ex}[P(\varepsilon_i |\phi_\lambda)]   \nonumber \\ 
& + & \beta \sum_{i>j} \rho_{ij}\sqrt{ \langle \delta\varepsilon_i^2 | \phi_\lambda\rangle \langle \delta\varepsilon_j^2 | \phi_\lambda\rangle} \, ,
\label{eq:corrsum}
\end{eqnarray}
where $\mu^{\rm ex}[P(\varepsilon_i |\phi_\lambda)]$ is described by Eq.~\ref{eq:gaussian}. 
The second summation in Eq.~\ref{eq:corrsum} can be rewritten as a sum over all nearest neighbor pairs $\sum_{i,i+1}$, the next nearest 
pairs $\sum_{i,i+2}$, etc. From the summation arranged in this fashion, we can then identify the effect of correlations at various
spatial length scales to the free energy $\mu^{\rm ex}[P(\varepsilon|\phi_\lambda)]$. Note that the present formulation precisely identifies the 
contributions solely due to the individual groups, namely the quantities $\mu^{\rm ex}[P(\varepsilon_i |\phi_\lambda)]$; we shall call this the zeroth-order
or self-contribution of the group $i$. For ease of presentation, when we speak of, say, $(i,i+2)$ correlation, we mean the correlation
between the binding energies of groups $i$ and $i+2$, respectively, with the solvent. 

We have pursued the above development for the long-range piece $\mu^{\rm ex}[P(\varepsilon|\phi_\lambda)]$, the solvation free energy of the conditioned
solute,  because in this case the binding energy distribution is well-behaved. Conceptually a decomposition similar to Eq.~\ref{eq:corrsum} can also be sought 
for $\mu^{\rm ex}$ (Eq.~\ref{eq:qc}), the net chemical potential of the solute. But in that case the functional form of the correlation contributions (which can be beyond linear-order) and even the individual contributions are, in general, difficult to ascertain. Our plan is to show that even for the idealized case of the conditioned solute, the
effect of near-neighbor correlations is non-trivial, and hence the increment in $\mu^{\rm ex}_n[P(\varepsilon|\phi_\lambda)]$ with $n$ 
is not solely a measure of the contribution of the individual group. That then implies that such a group transfer quantity will depend on the model system on which it was obtained. 
On this basis, given that $\mu^{\rm ex}[P(\varepsilon|\phi_\lambda)]$ is one component of $\mu^{\rm ex}$ and the greatly enhanced complexity of short-range solute-solvent effects, it is safe to conclude that the same conclusions hold for $\mu^{\rm ex}$ as well. 

\section{Results and Discussion}\label{sc:results}
\subsection{Solvation of the physical solute}
\begin{figure}[h!]
\centering
\includegraphics{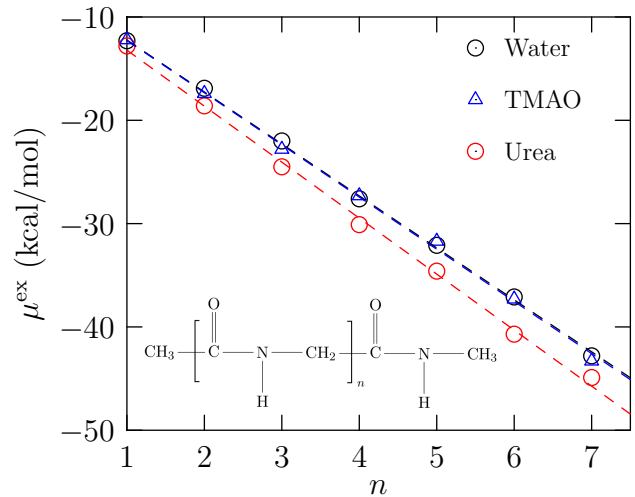}
\caption{The solvation free energy (Eq.~\ref{eq:qc}) versus $n$ for blocked (Gly)$_n$.}\label{fg:all}
\end{figure}

Fig.~\ref{fg:all} makes it clear that $\mu_n^{\rm ex}$ versus $n$ for blocked (Gly)$_n$ obeys a linear dependence. Similar linearity also holds for the chemical,
packing, and long-range contributions (Eq.~\ref{eq:qc}) individually. 
Per the constant increment method \cite{Auton:bio04}, we consider the slope of these curves as the contribution of an individual group to the free energy. These values are
collected in Table~\ref{tb:all} together with results for cGG. (Following established experimental approach \cite{Auton:bio04}, the value for cGG is divided by 2 to obtain
the value for one CH$_2$CONH group.) 
\begin{table}[h!]
\caption{Peptide group transfer free energies from vacuum to solvent obtained from the slope of $\mu^{\rm ex}_n$ versus $n$. Values for cGG have been scaled by 1/2.
Below each line for the model system studied, we present the transfer free energy values for transfer from water to the solution
under study. All values are in kcal/mol. Standard error of the mean is about 0.1~kcal/mol  (1$\sigma$).}\label{tb:all}
\centering
\begin{tabular}{lrrr} 
 &  Water & Urea & TMAO \\ \hline
 (Gly)$_n$ & $-5.0$ & $-5.4$ & $-5.0$ \\
                     &                          & $-0.4$ & $0$ \\ 
 cGG/2           & $-6.2$ & $-6.6$ & $-6.2$ \\ 
                    &                          & $-0.4$ & $0$ \\
\end{tabular}
\end{table}

The  water to the aqueous osmolyte transfer free energy agrees quite well for both the (Gly)$_n$ and cGG models. 
The urea concentration is about 8 M and assuming a linear dependence of transfer free energy on osmolyte concentration \cite{pace:jbc74,Hu:proteins2010}, we find that for 1 M
urea solution, the transfer free energy is $-50 \pm 13$~cal/mol, a value that is in good agreement with experimental estimates \cite{Auton:bio04}. 
We find a net zero transfer free energy to aqueous TMAO solution (4 M). This appears likely due to inadequacy in the
forcefield model for TMAO \cite{Hu:proteins2010,Canchi:jpcb12}.  

From Table~\ref{tb:all} we can note that the good agreement  in water to aqueous osmolyte solution transfer free energy 
masks the rather poor agreement in transfer free energies from vacuum to the respective solution. While it can be argued that water to osmolyte 
solution transfer is the most relevant experimentally, our results suggest that these small values arise from differences of substantially
large quantities. From the perspective of a physical theory, the vacuum to solution transfer quantities have the virtue of highlighting the role of
inter-group correlations transparently, and  in particular, Table~\ref{tb:all}
clearly shows that a peptide in (Gly)$_n$ is different from a peptide in cGG, assuming the validity of the divide-by-2 construct, itself suspect for 
reasons noted in Sec.~\ref{sc:csolute}. Based on analysis in Sec.~\ref{sc:csolute}, it seems plausible that in the water to osmolyte transfer free, 
inter-group correlations involving the (physical) solute cancel leaving a net change that is insensitive to the choice of the model system. 




\subsection{Solvation of the conditioned solute}\label{sc:csolute}

\begin{figure}[h!]
\centering
\includegraphics{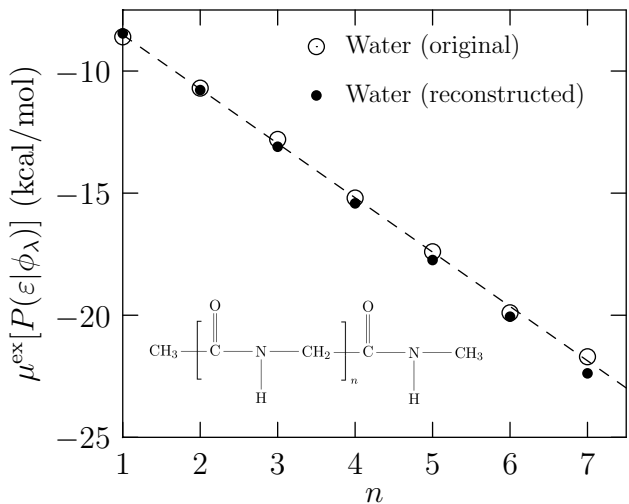}
\caption{Long-range contribution to the free energy of blocked (Gly)$_n$ in water. The open circles are the simulation results. The filled circles are based on using
the average values of the $i$ and $(i,i+1)$ correlation contributions from the (Gly)$_7$ chain (Table~\ref{tb:corr}) to reconstruct the free energy for all 
other $n$. The $(i,i+1)$ correlation contribution between groups 0 and 1 and between groups 7 and 8 for (Gly)$_7$ is used to model similar 
end-group correlations for all other $n$. Likewise, zeroth-order contributions from groups 0, 1, and 8 from (Gly)$_7$ are used for all $n$.}
\label{fg:lr}
\end{figure}

As before (Fig.~\ref{fg:all}), even the solvation free energy of the conditioned solute, $\mu_n^{\rm ex}[P(\varepsilon|\phi_\lambda)]$,
 depends linearly on $n$ (Fig.~\ref{fg:lr}). 
 For the analysis below, we exclusively focus on the vacuum to water transfer. 

For dissecting the correlation contributions to the slope, we focus on the internal groups of the peptide, as these are the ones changing with $n$. 
For the blocked (Gly)$_7$ model, group 0 is the methyl, group 1 is the CONHCH$_2$ group formed between the acetyl group and the N-terminus of the protein, and Group 8 is the
terminal CONHCH$_3$ group. The remaining six (6) CONHCH$_2$ groups are termed the internal groups. For the (Gly)$_3$ model,
per this convention there are two internal groups. In Table~\ref{tb:corr} we present the average contribution due to various orders of correlation between these internal groups. 
\begin{table}[h!]
\caption{Average values of correlation contributions of various orders per internal peptide unit. $i$ indicates that only contribution of the group with the solvent
is included (the first term on the right in Eq.~\ref{eq:corrsum}); $i,i+1$ indicates of 1$^{st}$ neighbor correlation and so on. 
All values are in kcal/mol. For reference, note that the slope of the $\mu_n^{\rm ex}[P(\varepsilon|\phi_\lambda)]$ versus $n$ curve (Fig.~\ref{fg:lr}) is $-2.23$~kcal/mol.}\label{tb:corr}
\centering
\begin{tabular}{lrr} 
		 &  (Gly)$_7$ & (Gly)$_3$ \\ \hline
$i$              &   $-1.53$  & $-1.58$ \\
$i,i+1$       &   $-0.79$  &  $-0.78$ \\
$i,i+2$       &  $0.25$  & --- \\
$i,i+3$       &  $-0.11$ &  --- \\ \cline{2-3}
Total           & $-2.18$  & $-2.37$ \\ \hline
\end{tabular}
\end{table}

Notice that the contribution from the zeroth order (or self term) ($i$, Table~\ref{tb:corr}) is fairly different from the 
slope of the $\mu_n^{\rm ex}[P(\varepsilon|\phi_\lambda)]$ versus $n$ curve. As Eq.~\ref{eq:corrsum} shows, this term --- the summands in the first term 
on the right of Eq.~\ref{eq:corrsum} ---  is also the one that can be rigorously identified as a contribution solely due to the group. Progressively 
including contributions from $(i,i+1)$, $(i,i+2)$, etc. correlations, we find a sum that is reasonably close to the slope of $\mu_n^{\rm ex}[P(\varepsilon|\phi_\lambda)]$ 
versus $n$. (The slight discrepancy between the sum computed in Table~\ref{tb:corr} and the slope arises because the linear fit
is not perfect.) Observe that the contribution from various orders of correlation are fairly similar for (Gly)$_7$ and (Gly)$_3$. Likewise, 
the correlation of the end groups with the internal groups are also fairly similar for these two models (data not shown),  implying insensitivity to chain length for the
correlations involving long-range interactions.  

It proves insightful to consider how well the average values of  various orders of correlation for (Gly)$_7$ model the free energy for all other chain 
lengths. Towards this end, we take the average value of the $(i,i+1)$ correlation contribution from Table~\ref{tb:corr}, the zeroth-order contributions for 
groups 0, 1, and 8,  and the $(i,i+1)$ contribution between groups 0 and 1 and between groups 7 and 8, and use Eq.~\ref{eq:corrsum} to compute the free energy for all $n$.  The good agreement for all $n$, including $n=1$ (which is all end-groups in our notation), reveals the  underlying uniformity of these self ($i$) and nearest neighbor ($i,i+1$) correlation in this model system. 

We can further appreciate the subtlety in these correlation contributions by identifying the electrostatic and dispersion contributions separately (Table~\ref{tb:elecvdw}).
For dispersion interactions, all orders of correlation beyond the zeroth-order tend to elevate $\mu^{\rm ex}_n[P(\varepsilon|\phi_\lambda)]$. This makes sense since
a favorable interaction of a water molecule with one center necessarily promotes a favorable interaction of that water with an adjacent center and vice versa. (For the 
dispersion interactions,  the relative orientation of the water molecule is irrelevant within the forcefield model.) 
\begin{table}[h!]
\caption{Net  contribution due to correlations of various orders for (Gly)$_7$. The column marked `All' gives the contribution with electrostatics and dispersion taken together. 
The columns marked `Elec' and 'vdW' give the contributions due to electrostatics and dispersion contributions separately. 
All values are in kcal/mol. For reference, the net free energy obtained by particle insertions is $-21.6$~kcal/mol.}\label{tb:elecvdw}
\centering
\begin{tabular}{lrrr} 
		 &  All & Elec & vdW \\ \hline
$i$              &   $-17.0$  & $3.0$  & $-20.6$\\
$i,i+1$       &   $-5.4$  &  $-6.4$ & $0.7$\\
$i,i+2$       &  $1.4$  & $1.2$ & $0.3$ \\
$i,i+3$       &  $-0.5$ &  $-0.6$ & $0.0$ \\ \cline{2-4}
Total           & $-21.5$  & $-2.6$ & $-19.6$ \\ \hline
\end{tabular}
\end{table}
For electrostatics, the relative orientations are important and near neighbor  interactions are anti-correlated: a favorable interaction of water with one site necessarily comes at the price of a favorable interaction with the adjacent site. For this same reason, the higher order electrostatic contributions tend to oscillate. Observe also that the sum of the electrostatic
and vdW contributions in Table~\ref{tb:elecvdw} is not precisely equal to the value when these are taken together. This arises because these individual contributions to the binding energy are themselves correlated, and separating them is only approximately true. Finally, consistent with Ref.\ \cite{paulaitis:corr10}, we find that the binding energy of the methyl end group (group 0) is anti-correlated with the binding energy of group 1 (data not shown); this result together with the data in Table~\ref{tb:elecvdw} thus suggests caution
in decoupling polar and non-polar group contributions, especially if these groups are adjacent in space. 

\begin{figure}[h!]
\centering
\includegraphics{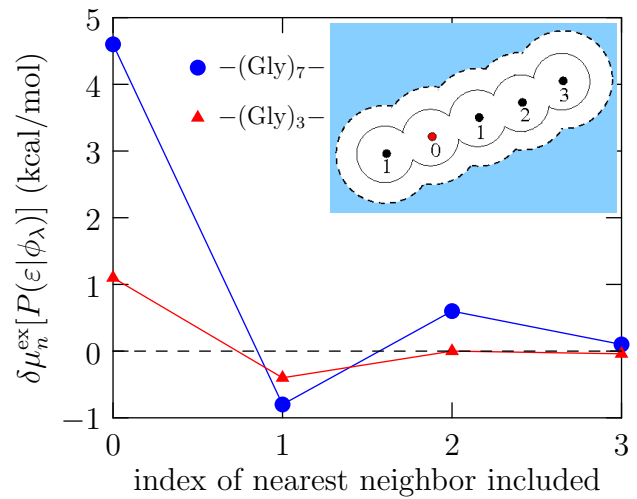}
\caption{$\delta\mu_n^{\rm ex}[P(\varepsilon|\phi_\lambda)]$ is the deviation of $\mu_n^{\rm ex}[P(\varepsilon|\phi_\lambda)]$ from the
net free energy (left hand side of Eq.~\ref{eq:corrsum}). The inset is a schematic to indicate order of nearest neighbors, relative to the group
labeled 0. Thus when no nearest neighbor is included, we sum only the individual group contributions (summands in the first term on the right
in Eq.~\ref{eq:corrsum}). Including the first neighbor means including all $i,i+1$ contributions to the free energy as well.}\label{fg:correrror}
\end{figure}

Fig.~\ref{fg:correrror} shows the deviation in the calculated $\mu_n^{\rm ex}[P(\varepsilon|\phi_\lambda)]$ relative to the net free energy (left hand
side of Eq.~\ref{eq:corrsum}) upon inclusion of increasing orders of correlation. It is evident that for (Gly)$_7$ correlations up to $i,i+3$ must be included to 
obtain a free energy that is converged. 

Table~\ref{tb:corrsolvs} compares the average values of the various orders of correlation in the solvation of (Gly)$_7$ in different solvents. Remarkably, we find
that all orders of correlation excluding the zeroth-order (self) contribution are identical. So at least in so far as the long-range interactions are concerned, the difference
in transfer free energy from water to the osmolyte solution can be entirely determined by the self-contribution, which is also the contribution that obeys group additivity. 
\begin{table}[h!]
\caption{Average values of various orders of correlation contributions per peptide unit for (Gly)$_7$ in different solvents. Rest as in Table~\ref{tb:corr}.}\label{tb:corrsolvs}
\centering
\begin{tabular}{lrrr} 
		 &  Water & Urea & TMAO  \\ \hline
$i$              &   $-1.53$  & $-1.97$  &  $-1.53$ \\
$i,i+1$       &   $-0.79$  &  $-0.79$ & $-0.79$ \\
$i,i+2$       &  $0.25$  & $0.25$ & $0.25$  \\
$i,i+3$       &  $-0.11$ & $-0.11$ & $-0.11$ \\ \cline{2-4}
Total           & $-2.18$  & $-2.62$ & $-2.18$ \\ \hline
\end{tabular}
\end{table}

The above analysis clearly shows that the incremental change in $\mu_n^{\rm ex}[P(\varepsilon|\phi_\lambda)]$ with respect to $n$ includes factors
beyond just the interaction of the added group with the solvent: \textbf{\textit{additivity does not imply independence}}.  Although it is not always stated
explicitly, group-additive transfer free energy contributions are treated as independent contributions. Our analysis shows this is, in general, invalid.

In contrast to our conclusion, the studies by Bolen and coworkers clearly show that the group-transfer model is capable of describing the $m$-value (the unfolding free energy in an osmolyte solution minus that in water) in a near quantitative fashion \cite{Auton:pnas05,Auton:bc11}. From the perspective of vacuum to solvent transfer, the $m$-value is a difference of difference involving four large transfer free energy contributions (two each for the unfolded and folded states of the protein, respectively) and some degree of cancellation of errors can be expected. But the level of agreement between calculated- and experimental-$m$ value \cite{Auton:pnas05,Auton:bc11} is remarkable and suggestive of some underlying physical regularity. Based on the similarity of higher-order correlation contributions for the conditioned solute in different solvents, and the observed linearity of even 
packing and chemistry contributions (data not shown), we suspect that in the $m$-value analysis, these higher order effects cancel, leaving only the self (or group-additive)
contribution intact.  Exploring this idea further is left for future studies.

\section{Concluding discussion}\label{sc:conclusions}

Group additivity has a hallowed place in chemistry; indeed it has even been referred to as the 4$^{\rm th}$ law of thermodynamics \cite{Dill:1997tg}. However, 
unlike a small molecular solute in the gas phase that is entirely characterized by strong, short-range interactions, in the treatment
of a many body system, such as a protein in a solvent, characterized by many different scales of energies, ranging from strong, short-range
interactions to relatively weak but fairly long-range interactions, group additive ideas must be considered with sufficient care. 

Even for an idealized solute with no short-range solute-solvent interaction, we find that the net solvation free energy of the solute comprises 
contributions due to the correlated interaction of the {solvent} with distinct groups in the solute. As is intuitively reasonable, the contribution of individual 
group-solvent interaction to the net free energy is the most dominant. However, the binding energy of a group $i$ with the solvent is correlated
with the binding energy of its neighbor $i\pm 1$, $i \pm 2$, etc.  These correlated fluctuations can either raise or lower the free energy of the solute, and
importantly, they can persist even for spatially distant groups. For the linear (Gly)$_7$, correlations persist up to $i,i+3$. 
A similar behavior, perhaps even longer-range of correlations, can be expected for a topographically complicated object. 
Further, the individual electrostatic and dispersion contributions to the binding energies are themselves correlated, making the identification of separate
free energy contributions due to polar and nonpolar groups problematic, especially if those groups are spatially close. 

Given that our analysis uncovers non-negligible effects of correlations even for an idealized solute with no direct solute-solvent contact, we must expect substantially more
involved correlation effects for a real solute, one that has an even more complicated short-range interaction with the solvent, and for solutes with formal charges,  
such as charged amino acid side-chain residues. In light of this identification, the physical basis for why such additive transfer models appear successful remains
to be explained. A plausible solution, suggested by this work, may rest in the similarity of the correlation contributions between different solutions. 


\section{Methods}
The simulation procedure closely follows our earlier work \cite{weber:jctc12} and only key differences are noted.  The peptides are modeled in the
extended configuration with the long axis aligned with the diagonal of the simulation cell and the center at the center of the simulation cell. (Initial configurations
were energy minimized with restraints to keep the peptide extended.) The peptide atoms are fixed in space throughout the simulation. The solvent was modeled by the TIP3P \cite{tip32,tip3mod} model and the CHARMM \cite{charmm}
forcefield with correction terms for dihedral angles \cite{cmap2}  was used for the protein.  A total of 2006 TIP3 molecules solvated the protein. Parameters for urea and TMAO were
obtained from Ref.~\cite{smith:urea03} and~\cite{Kast:tmao03}, respectively. A total of 449 urea molecules (for a molar concentration of about 8 M) and 195
TMAO molecules (for a molar concentration of about 4 M) were used. 
Unlike the earlier study \cite{weber:jctc12}, where the external field evacuated a spherical domain around the molecule, here we apply 
atom-centered fields to carve out a molecular cavity in the liquid;  the functional form of the field was as before (Eq.~4b, Ref.~\cite{weber:jctc12}). 
To build the field to its eventual range of $\lambda = 5$~{\AA}, we progressively apply the field, and for every unit {\AA} increment in the range, 
we compute the work done in applying the field using Gauss-Legendre quadratures. Five Gauss-points ($\lambda = 0, \pm (1/3) \sqrt{5-2\sqrt{10/7}}, \pm (1/3) \sqrt{5+2\sqrt{10/7}}$) are chosen for each unit {\AA}. At each Gauss-point, the system was simulated for 1~ns and the data from the last 0.5~ns used for analysis. (Excluding more data did not change the 
numerical value, indicating good convergence. Error analysis and error propagation was performed as before \cite{weber:jctc12}.) The starting configuration for each $\lambda$ point is obtained from the ending configuration of the
previous point in the chain of states. For the packing contributions, thus a total of 25 Gauss points span $\lambda \in [0,5]$.  For the chemistry contribution,
since solvent never enters $\lambda < 2.5$~{\AA}, we simulate $\lambda \in [2,5]$ for a total of 15 Gauss points. Separate calculations with a lower
order Gauss-Legendre quadrature and a trapezoidal rule (with $\lambda$ incremented in steps of 0.1~{\AA} \cite{weber:jctc12}) showed that results are very well 
converged with the five-point quadrature (data not shown). 

The long-range contribution $\mu^{\rm ex}_{n}[P(\varepsilon|\phi_\lambda]$ ($\lambda = 5$~{\AA}) was obtained by inserting the solute \cite{weber:jcp11}
 in a cavity (with atom-centered radius $\lambda = 5$~{\AA}). 1500 equally spaced cavity configurations were obtained from the last 0.375~ns of a 1~ns simulation at 
$\lambda = 5$~{\AA}. (The starting configuration for the $\lambda = 5$~{\AA} simulation was obtained from end point of the Gauss-Legendre procedure
as indicated above.) We also did solute extraction calculations \cite{asthagiri:jacs07} in a like fashion, with 5000 binding energy values obtained over 0.5~ns of simulation. 
Confirming the Gaussian distribution of binding energies, both procedures gave free energies to within 0.1~kcal/mol of each other (data not shown). 
The binding energies for the correlation analysis were obtained from the solute extraction procedure. 

Cyclic-diglycine was built and optimized using the Gaussian (G09) quantum chemistry package \cite{g09}. For consistency with the (Gly)$_n$ simulations, the 
partial charges and Lennard-Jones interaction parameters were obtained from the backbone atoms of the CHARMM forcefield.

\begin{acknowledgements}
We thank George Rose for helpful discussions and critical insights. This research used resources of the National Energy Research Scientific Computing Center, which is supported by the Office of Science of the U.S. Department of Energy under Contract No. DE- AC02-05CH11231.
\end{acknowledgements}

%

\end{document}